\documentstyle[amssymb,preprint,aps]{revtex}

\begin{document}
\author{S. Bruce}
\address{Departamento de Fisica, Universidad de Concepcion,\\
Casilla 160-C, Concepcion, Chile\\
E-mail: sbruce@udec.cl}
\title{Comments on discrete time in quantum mechanics}
\date{March, 2001}
\maketitle

\begin{abstract}
The possibility that time can be regarded as a discrete parameter is
re-examined. We study the dynamics of the free particle and find in some
cases superluminal propagation.



\bigskip 
\end{abstract}

\bigskip \newpage 

\section{Introduction}

The goal of any physical theory is to set up a mathematical model which
enables us to correlate some empirical phenomena. A physical theory is
considered satisfactory if we can make with it quantitative predictions of
physical data. Such data usually involve the measurements of certain
quantities which are expressed in a system of fundamental units. The choice
of these units is to some extent arbitrary with respect to magnitude as well
as with respect to kind. Such a choice of units will therefore be guided
primarily by considerations of convenience.

A great simplification is introduced in quantum mechanics if we use as the
fundamental units certain physical quantities which are constants of nature.
Two constants of this sort are $c$, the speed of light in vacuum, and $\hbar 
$, Planck's bar constant. We usually choose these constants as two of the
fundamental units of our systems. As the third unit we use the second or
centimeter as the conventional and arbitrary unit of time or length. Here we
shall mention that this choice is guided by the fact that the theory under
discussion arises from an intimate relation of special relativity
characterized by the constant $c$ and quantum mechanics characterized by $%
\hbar $. To our knowledge, at present there exists no theory which involves
in its fundamental laws either a universal time or a universal length, which
would make a natural choice of the third basic unit. The need for such a
theory involving a fundamental time (length) has been the subject of much
speculation in the past and present but it seems safe to say that we are far
from understanding the role of such a unit in existing theories.

Throughout the development of quantum mechanics, time always appears as a
continuous parameter. Take the example of a nonrelativistic particle. In
Feynman's path integration formulation, the probability amplitude for the
particle to be at the position ${\bf q}(t_{i})$ and at ${\bf q}(t_{f})$ is
given by the amplitude sum over all paths ${\bf q}(t)$ connecting ${\bf q}%
(t_{i})$ and ${\bf q}(t_{f})$, apart from a normalization constant. Clearly,
the position of the particle ${\bf q}$ is not treated on the same basis as
the (real) time $t$: at a given time the path integration can be viewed as
that over the whole range of eigenvalues of the position operator. This then
underlies the familiar difference between ${\bf q}$ as an operator and $t$
as a parameter.

In fact this asymmetry can be made out in classical mechanics. The classical
trajectory of a particle is determined by the extremity of the action, which
is a functional of ${\bf q}(t)$. While ${\bf q}$ is the dynamical variable, $%
t$ appears only as a continuous parameter. By setting the variational
derivative, we obtain the usual Lagrange equation of motion, whose solution
gives the classical path. In relativistic quantum field theory, space and
time have to be treated symmetrically due to Lorentz invariance. The usual
approach is to regard $q_{i}$ and $t$ all as parameters; the operators are
then field variables.

The purpose here is not to replace a continuous dynamical evolution
parameter with a discrete parameter. Our interest is in the construct of a
self-consistent discrete complex-time quantum mechanics with well specified
equations of motion. This is motivated by the notion that at some small
scale, time is really discrete. This has echoes in theories such as
Relativistic quantum mechanics with a time associated to the electron's
Compton wavelength $\left( 10^{-22}\text{s}\right) ,$ and string theory,
where the Planck time $(10^{-43}$s$)$ sets a scale at which conventional
notions of space and time break down.

There are various circumstances in physics where it is convenient or
necessary to replace the continuous time (temporal evolution) parameter with
a discrete parameter. There have been various attempts lo construct
classical and quantum mechanical theories based on this notion, such as the
work of Caldirola \cite{CA} and Lee \cite{LE}. The work of Yamamoto et al 
\cite{YA}, Hashimoto et al \cite{HA}, Klimek \cite{KL}, Jaroszkiewicz and
Norton \cite{JA}, and Milburn \cite{MI,BU}, show that the subject continues
to receive attention.

The underlying postulate is that on sufficiently short time steps the system
does not develop continuously under a mixture of unitary and nonunitary
evolution but rather in a sequence of {\it identical} transformations. The
inverse of this time step is the mean frequency of the steps, $\delta t$,
which turns into an expansion parameter. If the time step is large enough,
the evolution appear approximately continuous on laboratory time scales. To
zeroth order the Schr\"{o}dinger equation is recovered.

One feature of this model is that constants of the motion remain constants
of the motion and thus stationary states remain stationary states. Whether
or not these consequences are observable depends on the size of $\delta t.$

In the following we wish to explore some alternative possibilities. First,
in place of treating time as a real parameter, we may consider time as a
continuous {\it complex} parameter (analytical continuation formulation).
Second, time can be treated as a {\it discrete }complex parameter (discrete
complex-time formulation). As we shall see, both possibilities can be
realized. The result is that in this new formalism our usual idea of
continuous time structure will appear only as an approximation.

\section{Continuous time evolution}

Let us consider a quantum system whose time evolution is given by the
complex time propagator 
\begin{equation}
U\left( s\right) =\exp \left( sH\right) ,  \label{pro}
\end{equation}
where $s=\left( -i/\hbar \right) \left( t+iv\right) $ with $t$ and $\nu $
real parameters. In the above the Hamiltonian $H$ is assumed to be Hermitian
and time-independent. In a particular physical system we look for a complete
set of commuting observables $\hat{\alpha}$. We then can take their
simultaneous eigenkets as basic kets: $\left| \Psi _{\alpha }\left( s\right)
\right\rangle ,$ $\alpha =(\alpha _{1},\alpha _{2},...)$, where $\alpha _{i}$
is the eigenvalue of the observable $\hat{\alpha}_{i}.$ The Schr\"{o}dinger
equation for the system is then 
\begin{equation}
H\left| \Psi _{\alpha }\left( s\right) \right\rangle =\frac{d}{ds}\left|
\Psi _{\alpha }\left( s\right) \right\rangle =E_{\alpha }\left| \Psi
_{\alpha }\left( s\right) \right\rangle .
\end{equation}
The reason for looking at the propagator (\ref{pro}) will be clearer as we
go along. First let us note that we can write down the formula for it at
once: 
\begin{equation}
U\left( s\right) =\sum_{\alpha }\left| \Psi _{\alpha }\left( s\right)
\right\rangle \left\langle \Psi _{\alpha }\left( s\right) \right| \exp
\left( sE_{\alpha }\right) .
\end{equation}
The main point to note is that even though the time is now complex, the
eigenvalues and eigenfuctions that enter into the formula for $U\left(
s\right) $ are the usual ones. Conversely, if we knew $U\left( s\right) $,
we could extract the former.

The Hamiltonian is assumed to be represented by a self adjoint operator.
According to the basic principles of QM one defines a Hilbert space ${\cal H}
$ for each QM system. Every measurable quantity or ``observable'' is
represented by a self adjoint operator. The state of the system at time $s $
is given by a vector $\left| \Psi \left( s\right) \right\rangle \in {\cal H}$
which is analytic in the complex plane defined by $s.$ Note that 
\begin{equation}
\frac{d}{ds}\left| \Psi \left( s\right) \right\rangle =i\hbar \frac{\partial 
}{\partial t}\left| \Psi \left( s\right) \right\rangle =\hbar \frac{\partial 
}{\partial \upsilon }\left| \Psi \left( s\right) \right\rangle .
\end{equation}
The states $\Psi $ are then normalizable, i.e., 
\begin{equation}
\left\langle \Psi \left( s\right) |\Psi \left( s\right) \right\rangle =\int_{%
{\cal V}}\Psi ^{\dagger }\left( {\bf q},s\right) \Psi \left( {\bf q}%
,s\right) d^{3}q\leqq 1,\qquad \forall \ s,
\end{equation}
where ${\cal V}$ is the volume where the system is contained. The ``%
\mbox{$<$}%
'' sign stands for possible intrinsic loss of information in the quantum
system. Given a basis $\left\{ \left| \Psi _{j}\left( s\right) \right\rangle
\right\} $ of ${\cal H}$ a self-adjoint operator is defined as satisfying 
\begin{equation}
\left\langle \Psi _{i}\left( s\right) \right| {\cal O}\left| \Psi _{j}\left(
s\right) \right\rangle =\left\langle \Psi _{j}\left( s\right) \right| {\cal O%
}\left| \Psi _{i}\left( s\right) \right\rangle ^{\ast }.  \label{SA}
\end{equation}
The stationary states can be written in the form 
\begin{equation}
\left\langle {\bf q}|\Psi _{E_{\alpha }}\left( s\right) \right\rangle =\Psi
_{E_{\alpha }}\left( {\bf q},s\right) =\exp \left( sE_{\alpha }\right)
u_{E_{\alpha }}\left( {\bf q}\right) ,
\end{equation}
with 
\begin{equation}
Hu_{E_{\alpha }}=Eu_{E_{\alpha }}\ .
\end{equation}
Notice that the (conformal) mapping $s\rightarrow $ $\exp \left( sE_{\alpha
}\right) $, which has no zeros and no singularities in the entire complex
plane, turns out to posses an essential singularity at infinity.

\section{General evolution}

The feature of quantum mechanics that most distinguishes it from classical
mechanics is the coherent superposition of distinct physical states. This
feature is at the heart of the less intuitive aspects of the theory. It is
the basis for the concern about measurement in quantum mechanics, and it is
the explanation for the nonappearance of chaos in systems that classically
would be chaotic. Apparently, however, the superposition principle does not
operate on macroscopic scales, although nothing in the present formulation
of quantum mechanics would indicate this.

We now consider an ad hoc time distribution $0,s_{1},s_{2},...,s_{N},$ in
the complex plane where $s_{i}-s_{j}=\delta s$ is a fundamental time
interval. Thus in the Schroedinger equation we need to introduce a {\it %
discrete derivative} associated with the given time distribution, namely 
\begin{equation}
H_{D}\left| \Psi _{\alpha }\left( s\right) \right\rangle =\frac{\delta
_{\lambda }}{\delta s}\left| \Psi _{\alpha }\left( s\right) \right\rangle ,
\end{equation}
with $H_{D}$ in the $s$ representation, where $s$ is a given $s_{j}$, and 
\begin{equation}
\frac{\delta _{\lambda }}{\delta s}\left| \Psi _{\alpha }\left( s\right)
\right\rangle \equiv \lambda ^{-1}\left( s,\delta s\right) \left( \left|
\Psi _{\alpha }\left( s+\delta s\right) \right\rangle -\left| \Psi \left(
s+\delta s-\lambda \left( s,\delta s\right) \right) \right\rangle \right)
\label{D}
\end{equation}
where $\lambda $ is an {\it holomorphic} function of $s$ in the whole
complex plane and of order $\delta s$, with $\delta s$ a given finite
difference in the complex {\it time}-plane. This is to be interpreted as a
more general time evolution being the continuous evolution a limit case.

In the above 
\begin{equation}
s=-\frac{i}{\hbar }\left( t+iv\right) ,\qquad \delta s\equiv -\frac{i}{\hbar 
}\left( \delta t+i\delta v\right) ,
\end{equation}
and $\left| \Psi _{\alpha }\left( s\right) \right\rangle $ is {\it analytic }%
at $s$, therefore it can be expanded in a Laurent series. Thus 
\begin{equation}
\frac{\delta _{\lambda }}{\delta s}\left| \Psi _{\alpha }\left( s\right)
\right\rangle =\frac{d}{ds}\left| \Psi _{\alpha }\left( s\right)
\right\rangle +\sum_{n=2}^{\infty }\frac{\lambda _{s}^{n-1}}{n!}\frac{d^{n}}{%
ds^{n}}\left| \Psi _{\alpha }\left( s\right) \right\rangle ,
\end{equation}
i.e., 
\begin{equation}
H_{D}=\frac{2}{\lambda _{s}}\exp \left( \left( \delta s-\frac{\lambda _{s}}{2%
}\right) H\right) \sinh \left( \frac{\lambda _{s}}{2}H\right) ,  \label{Htil}
\end{equation}
where $\lambda _{s}\equiv \lambda \left( s,\delta s\right) .$ For $\lambda
_{s}$ a constant function of $s,$ the stationary states are of the form 
\begin{equation}
\Psi _{E_{D}}\left( {\bf q},s\right) =e^{s\epsilon }u_{E}\left( {\bf q}%
\right) =e^{\frac{1}{\hbar }\left( t\epsilon _{I}+v\epsilon _{R}\right) }e^{-%
\frac{i}{\hbar }\left( t\epsilon _{R}-v\epsilon _{I}\right) }u_{E}\left( 
{\bf q}\right) ,
\end{equation}
with $E$ the eigenvalues of $H$ and $\epsilon \equiv \epsilon _{R}+i\epsilon
_{I}.$ In other words, we assume that $H$ and $H_{D}$ act on the same
Hilbert space. Furthermore if 
\begin{equation}
H_{D}u_{E}\left( {\bf q}\right) =E_{D}u_{E}\left( {\bf q}\right) ,
\end{equation}
we find that 
\begin{equation}
E_{D}=\frac{2}{\lambda _{s}}\exp \left( \left( \delta s-\frac{\lambda _{s}}{2%
}\right) E\right) \sinh \left( \frac{\lambda _{s}}{2}E\right) .  \label{ED}
\end{equation}
We must require some basic physical conditions upon the energy eigenvalues.
Particularly, they must take real values, which means that $%
\mathop{\rm Im}%
E_{D}=0$, a condition we have to impose on (\ref{ED}).

\section{Examples}

We shall consider two particular cases where the Hamiltonian $H_{D}$ is
Hermitian.

Case a) $\lambda _{s}=$ $\delta s=\left( 1/\hbar \right) \tau _{1}$, with $%
\tau _{1}$ a finite time element. This case corresponds to an intrinsic loss
of information. The discrete derivative is 
\begin{equation}
H_{D}=\frac{\delta _{\lambda }}{\delta s}=\frac{2\hbar }{\tau _{1}}\exp
\left( \frac{\tau _{1}}{2}i\frac{\partial }{\partial t}\right) \sinh \left( 
\frac{i\tau _{1}}{2}\frac{\partial }{\partial t}\right) =\frac{\hbar }{\tau
_{1}}\left( \exp \left( \tau _{1}i\frac{\partial }{\partial t}\right)
-I\right) .  \label{HD}
\end{equation}
Thus 
\begin{equation}
H_{D}\left| \Psi _{\alpha }\left( s\right) \right\rangle =\frac{\delta
_{\lambda }}{\delta s}\left| \Psi _{\alpha }\left( s\right) \right\rangle
=\sum_{n=1}^{\infty }\frac{\left( \delta s\right) ^{n-1}}{n!}\frac{d^{n}}{%
ds^{n}}\left| \Psi _{\alpha }\left( s\right) \right\rangle .
\end{equation}

To determine the commutator $\left[ H_{D},s\right] $ we evaluate this
commutator operating on $\Psi \left( s\right) $, i.e., 
\begin{equation}
\left[ H_{D},s\right] =\exp \left( \frac{\tau _{1}}{\hbar }H\right) =i\hbar
\left( I+\frac{\tau _{1}}{\hbar }H_{D}\right) ,
\end{equation}
which involves a modification to the standard time-energy commutation
relation.

Next let us consider a relativistic spin-0 free particle. The Hamiltonian
becomes 
\begin{equation}
H_{D}=\frac{\hbar }{\tau _{1}}\left( \exp \left( \frac{\tau _{1}}{\hbar }%
H\right) -I\right) ,
\end{equation}
where $H=\sqrt{c^{2}{\bf p}^{2}+m^{2}c^{4}},$ ${\bf p=}-i\hbar {\bf \nabla }$%
. Therefore 
\begin{equation}
\frac{\delta \hat{q}_{j}}{\delta s}=\frac{i}{\hbar }\left[ H_{D},\hat{q}_{j}%
\right] =H^{-1}\exp \left( \frac{\tau _{1}}{\hbar }H\right) c^{2}p_{j},
\end{equation}
where 
\begin{equation}
\hat{q}_{j}\equiv q_{j},\qquad \hat{p}_{j}\equiv \frac{H_{D}}{c^{2}}\frac{%
\delta \hat{q}_{j}}{\delta s}=H^{-1}H_{D}\exp \left( \frac{\tau _{1}}{\hbar }%
H\right) p_{j}  \label{HD2}
\end{equation}
are canonical conjugate coordinates, with $p_{j}=-i\hbar \partial /\partial
q_{j}.$ Furthermore 
\begin{equation}
\frac{\delta \hat{q}_{j}}{\delta s}=H^{-1}\exp \left( \frac{\tau _{1}}{\hbar 
}H\right) c^{2}p_{j},
\end{equation}
corresponds to the {\it group velocity} of the quantum waves. Thus from (\ref
{HD2}) we get 
\begin{eqnarray}
\left[ \hat{q}_{i},\hat{q}_{j}\right] &=&\left[ \hat{p}_{i},\hat{p}_{j}%
\right] =0, \\
\left[ \hat{q}_{i},\hat{p}_{j}\right] &=&i\hbar \left( I+\frac{\tau _{1}}{%
\hbar }H_{D}\right) \left( H^{-1}H_{D}\delta _{ij}+\left( I-\left( I-\frac{%
2\tau _{1}H}{\hbar }\right) H_{D}H^{-1}\right) c^{2}H^{-2}p_{i}p_{j}\right) 
\nonumber \\
&=&i\hbar \left( I+\frac{3}{2}\frac{\tau _{1}}{\hbar }H_{D}\right) \delta
_{ij}+i\frac{3}{2}\frac{\tau _{1}c^{2}}{\hbar }H_{D}^{-1}p_{i}p_{j}+O\left(
\left( \frac{\tau _{1}}{\hbar }\right) ^{2}\right) .  \nonumber
\end{eqnarray}
Let us now postulate the existence of a relativistic invariant mass $M$ in
this context. To this end we impose the condition 
\begin{equation}
E_{D}^{2}-c^{2}{\bf \hat{p}}^{2}=E_{D}^{2}\left( 1-\exp \left( \frac{2\tau
_{1}E}{\hbar }\right) \left( 1-\left( \frac{mc^{2}}{E}\right) ^{2}\right)
\right) =M^{2}c^{4}.
\end{equation}
For ${\bf p}={\bf 0}$ we get 
\begin{equation}
M=\frac{\hbar }{\tau _{1}c^{2}}\left( \exp \left( \frac{\tau _{1}}{\hbar }%
mc^{2}\right) -1\right) =m+\frac{\tau _{1}}{2\hbar }m^{2}c^{2}+O\left(
\left( \frac{\tau _{1}}{\hbar }\right) ^{2}\right) ,
\end{equation}
which represents a shift in value of the inertial mass $m.$

For a massless particle 
\begin{equation}
E_{D}=\frac{\hbar }{\tau _{1}}\left( \exp \left( \frac{\tau _{1}c}{\hbar }%
\left| {\bf p}\right| \right) -1\right) ,\qquad v_{j}\left( \tau _{1}\right)
=c\exp \left( \frac{\tau _{1}c}{\hbar }\left| {\bf p}\right| \right) \frac{%
p_{j}}{\left| {\bf p}\right| }.
\end{equation}
As it is apparent, even for very small time step $\tau _{1},$ this model
predicts the possibilities of superluminal propagation$\ $provided that $%
\left| {\bf p}\right| \gtrapprox 0.$

Case b) We choose $\lambda _{s}=2\delta s,${\it \ }with $\delta s=\left(
-i/\hbar \right) \tau _{0}$ the step time element. The generalized
Schr\"{o}dinger equation is 
\begin{equation}
H_{D}\left| \Psi _{\alpha }\left( s\right) \right\rangle =\frac{\delta
_{\lambda }}{\delta s}\left| \Psi _{\alpha }\left( s\right) \right\rangle ,
\end{equation}
with 
\begin{equation}
\frac{\delta _{\lambda }}{\delta s}\left| \Psi _{\alpha }\left( s\right)
\right\rangle =\frac{1}{2\delta s}\left( \left| \Psi \left( s+\delta
s\right) \right\rangle -\left| \Psi \left( s-\delta s\right) \right\rangle
\right) ,
\end{equation}
the so-called {\it symmetric derivative. }Thus this is an instance of
unitary evolution. The Hamiltonian takes the form \cite{CA} 
\begin{equation}
H_{D}=\frac{\hbar }{\tau _{0}}\sin \left( \frac{\tau _{0}}{\hbar }H\right) =%
\frac{\hbar }{\tau _{0}}\sin \left( i\tau _{0}\frac{\partial }{\partial t}%
\right) .  \label{HS}
\end{equation}
Therefore the velocity operator is now given by 
\begin{equation}
\frac{\delta \hat{q}_{j}}{\delta s}=\frac{i}{\hbar }\left[ H_{D},\hat{q}_{j}%
\right] =H^{-1}\cos \left( \frac{\tau _{0}}{\hbar }H\right) c^{2}p_{j}.
\label{VE}
\end{equation}
The position and momentum operators are then 
\begin{equation}
\hat{q}_{j}\equiv q_{j},\text{\qquad }\hat{p}_{j}\equiv \frac{H_{D}}{c^{2}}%
\frac{\delta \hat{q}_{j}}{\delta s}=\frac{\sin \left( 2\tau _{0}H/\hbar
\right) }{2\tau _{0}H/\hbar }p_{j},  \label{CAN}
\end{equation}
with $p_{j}=-i\hbar \partial /\partial q_{j}$ as before$.$ Thus from (\ref
{CAN}) we get 
\begin{eqnarray}
\left[ \hat{q}_{i},\hat{q}_{j}\right] &=&\left[ \hat{p}_{i},\hat{p}_{j}%
\right] =0, \\
\left[ \hat{q}_{i},\hat{p}_{j}\right] &=&\delta _{ij}i\hbar \frac{\sin
\left( 2\tau _{0}E/\hbar \right) }{2\tau _{0}E/\hbar }+i\hbar \left( \cos
\left( 2\tau _{0}E/\hbar \right) -\frac{\sin \left( 2\tau _{0}E/\hbar
\right) }{2\tau _{0}E/\hbar }\right) \frac{c^{2}p_{i}p_{j}}{E^{2}}  \nonumber
\\
&=&\delta _{ij}i\hbar \left( I-\frac{2}{3}\left( \frac{\tau _{0}}{\hbar }%
\right) ^{2}H_{D}^{2}\right) -i\hbar \frac{4}{3}\left( \frac{\tau _{0}}{%
\hbar }\right) ^{2}c^{2}p_{i}p_{j}+O\left( \left( \frac{\tau _{0}}{\hbar }%
\right) ^{3}\right) .  \nonumber
\end{eqnarray}

From (\ref{HS}) and (\ref{VE}) we find that 
\begin{equation}
E_{D}=\frac{\hbar }{\tau _{0}}\sin \left( \frac{\tau _{0}}{\hbar }E\right)
,\qquad v_{j}\left( \tau _{0}\right) =c\cos \left( \frac{\tau _{0}}{\hbar }%
E\right) \frac{cp_{j}}{E}.  \label{f}
\end{equation}
The first Eq.(\ref{f}) tells us that a maximum value of the energy exists: $%
E_{D}=\hbar /\tau _{0}.$ If, for instance, $\tau _{0}$ is taken to be the
time associated with the Compton wavelength of the particle, $\tau
_{0}=\hbar /mc^{2},$ then 
\begin{equation}
E_{D}=mc^{2}\sin \left( \frac{E}{mc^{2}}\right) .
\end{equation}
Therefore, $E_{D}$ reaches its maximum value $\left( mc^{2}\right) $ for
relativistic values of $E$ $\left( \backsim \pi mc/2\gtrsim mc\right) ,$
i.e., when creation and annihilation of particles take place. In such a case
we must call for a (quantum) field theory treatment.

\section{Final comments}

To conclude, on sufficiently small time scales we conjecture that the system
evolves by a sequence of time-like steps generated by the Hamiltonian. The
Schr\"{o}dinger equation can be obtained to $n$th order in the expansion
(complex) parameter $\delta s.$ This time-like discretization involves a
modification both to the standard time-energy commutation relation and to
the $q,p$ canonical commutation relations. Particularly, for the case a)
even for small time step $\delta s,$ the model predicts the possibility of
superluminal velocities, for instance,$\ $for a massless particle with
nonvanishing momentum.

If the physical evolution time scale is much larger than the discrete time
scale then the evolution resembles to a quantum stochastic process. This
process could be studied using the quantum version of the Ito and
Stratonovich stochastic calculus \cite{ITO}.

\end{document}